\documentclass[twocolumn,twoside,slac_two]{revtex4}
\usepackage{graphicx}
\usepackage{fancyhdr}
\begin{document}

\title{VLBI and LAT 2-year results for the Bologna Complete Sample.}

\author{E. Liuzzo, G. Giovannini}
\affiliation{Istituto di Radioastronomia, Via P. Gobetti 101, 40129 Bologna (Italy)}
\affiliation{Dipartimento di Astronomia, Universit\`a di Bologna, Via Ranzani 1, 40127 Bologna (Italy)}

\author{M. Giroletti}
\affiliation{Istituto di Radioastronomia, Via P. Gobetti 101, 40129 Bologna (Italy)}

\begin{abstract}
The statistical analysis of parsec scale region of radio galaxies is crucial to obtain information on the nature of their central engine. To this purpose, we defined and observed the Bologna Complete Sample (BCS) which is unbiased with respect to the orientation of the nuclear relativistic jet being selected from low-frequency samples. The BCS is a complete sample of 94 nearby (z$<$0.1) radio galaxies that are well studied targets with literature kiloparsec data. For all of them, we collected parsec scale
information asking new VLBI (VLBA and EVN) observations. Statistical results on their properties in radio band are presented. From the estimates of the Doppler factor and viewing angles, we discuss the connection with the available gamma-ray data.
Finally, we show how future observations with Fermi could reveal new important detections of some of the BCS sources.
\end{abstract}

\maketitle

\thispagestyle{fancy}

\section{INTRODUCTION.}

The analysis of the parsec scale region of radio galaxies is crucial to obtaining information on the nature of their central engine. To statistically study properties of different classes of sources, it is necessary to define and observe a sample that is free of selection effects. With this aim, we began a project to observe a complete sample of radio galaxies selected from low frequency surveys, unbiased with respect to the orientation of the nuclear relativistic jet. In fact, sources from low frequency samples are dominated by their extended and unbeamed (isotropic) emission.

We selected sources in the B2 Catalogue and the Third Cambridge Revised Catalogue (3CR) , naming this ``the Bologna Complete Sample (BCS)''. For details see also \cite{gio90,gio01,gio05,liu09b}.\\
The sample is composed by 94 sources answering at the following criteria:
\begin{itemize}
\item flux density $<$ 0.25 Jy at 408 MHz for the B2 sources and greater than 10 Jy at 178 MHz for the 3CR sources (\cite{fe84})
\item declination $>$ $10^{\circ}$; 
\item  Galactic latitude $\vert b\vert>15^{\circ}$;
\item redshift z$<$0.1. 
  \end{itemize}

To properly map their nuclear emission, we asked and obtained observations with the standard VLBI technique that allows us a typically resolution of $\sim$ 2 mas which corresponds at 1.7 pc at z=0.09.

Radio and gamma-ray emissions in Active Galactic Nuclei are both related to the presence of relativistic particles in jets. Interestingly, the vast majority of identified extragalactic gamma-ray sources in the third EGRET catalog belong to the blazar class and 96/181 of the high galactic latitude ($∣b∣ >$ 10$^{\circ}$) sources remained unidentified (3EG, \cite{ha99}). With the advent of the Fermi Large Area Telescope (LAT), and thanks to its large sensitivity up to several GeV, it is being possible to associate counterparts and study the multi-wavelength properties of a large number of gamma-ray sources. Several observational results are changing our understanding of these phenomena. 
We used our study on the radio parsec scale morfologies and jet properties of BCS sources, together with the results of the LAT 2-year Catalog (2FGL, \cite{fermi2}), to investigate the relationship between gamma-ray and radio emission for a complete sample of nearby radiogalaxies.

We assume here a Hubble constant H$_{0}$= 70 km s$^{-1}$ Mpc$^{-1 }$, $\Omega_{\rm M}=0.3$ and $\Omega_{\lambda}=0.7$.

\section{RADIO DATA}

We asked and obtained Very Long Baseline Array (VLBA) observing time at 5 GHz to produce high-resolution images for sources of the BCS with an estimated nuclear flux density S$_{c, 5GHz}$ at arcsecond resolution $>$5 mJy. For fainter targets, data with VLBA and EVN at 1.6 GHz were carried out. Images for the 80 $\%$ (76/94) of the our sample are yet published by us (\cite{gio05, liu09b}). We are analysing the last observational blocks (18 sources) in L band (\cite{liu12}). As in most cases, the VLBA target is not sufficiently strong for fringe-fitting and self-calibration, we used the phase referencing technique using a 4 min duty cycle (2.5 min on source, 1.5 min on phase calibrator). In addition, a relatively long integration time for each source has been used to obtain good (u,v)-coverage, necessary to properly map complex faint structures. The noise level was estimated from the final images and it is typically $\sim$ 0.1 mJy/beam.

\subsection{Source morphology}

The detection rate of our VLBI data analysed up to now is high: only 3 sources out of 76 (4$\%$) have not been detected, even though we observed sources with an arcsecond core flux density as low as 5 mJy at 5 GHz. This result confirms the presence of compact radio nuclei at the center of radio galaxies.

For all sources, high-quality images at arcsecond resolution obtained with the Very Large Array (VLA) of the NRAO are available in literature, allowing a detailed study of the large-scale structure. Looking at the parsec and kpc scale structure of BCS radiogalaxies, we found that:

\begin{figure*} [th!]
\begin{center}
\includegraphics[width=110mm]{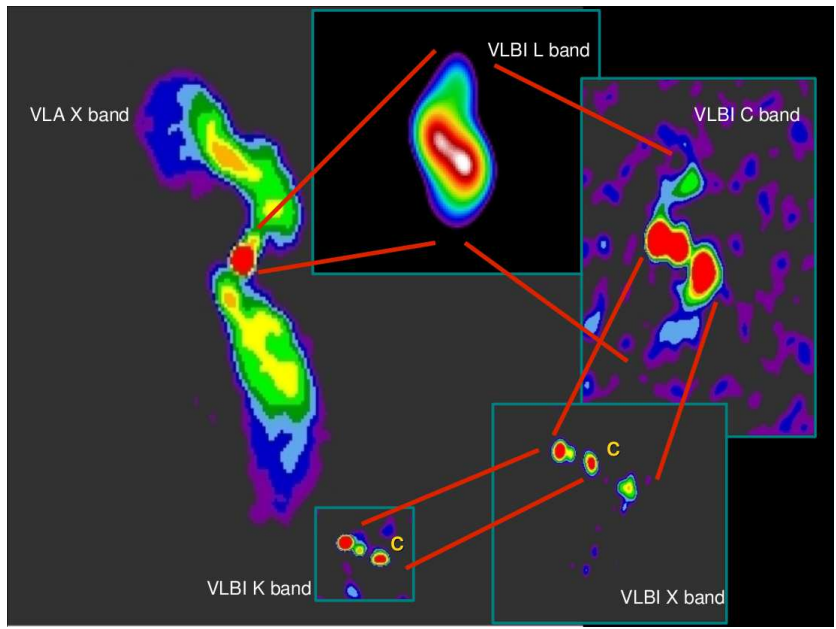}
\end{center}
\caption{Clockwise from left to right, zooming from kiloparsec to mas scale radiostructure of 4C 26.42 (\cite{liu09b}): color maps of VLA X band, VLBI L band, VLBI C band, VLBI X band and VLBI K band data. (C) indicates the core component.  }
\label{1346}
\end{figure*}

\begin{itemize}
\item  According to the kiloparsec scale morphology, the sample contains 65 FRI radio galaxies, 16 FRII sources, and 13 compact sources. 
\item  As expected in sources with relativistic parsec-scale jets, the one-sided jet morphology is the predominant structure present in our VLBI images, however $22\%$ of the observed sources show evidence of a two-sided strucure. This result is in agreement with a random orientation and a high jet velocity ( $\beta \sim 0.9$).
\item Two sources (4C26.42, \cite{liu09b}, Fig.~\ref{1346} and 3C 310 \cite{liu09b}) with a Z-shaped structure on the parsec-scale. The presence of this symmetric and large change in the jet direction so near the core cannot be due to projection or relativistic effects. The large symmetry present in these sources imply that are oriented close to the plane of the sky, moreover a change in the jet velocity and/or orientation would produce a change in the Doppler factor that should be revealed by the observations of the jet properties. Based on these considerations on the source morfology we suggest the presence of low velocity jets in these two peculiar radio sources.
\item In most cases (62 $\%$), the parsec and the kiloparsec scale jet structures are aligned and the main jet is always on the same side with respect to the nuclear emission. This confirms the idea that the large bends present in some BL Lacs sources are amplified by the small jet orientation angle with respect to the line-of-sight (\cite{gi04}).
\item In 62$\%$ of the sources, there is good agreement between the arcsecond-scale and the VLBI correlated flux density. For the other 38$\%$ of the sources, at the milliarcsecond scale more than 30$\%$ of the arcsecond core flux density is missing. This suggests the presence of variability, or of a significant sub-kiloparsec-scale structure, which will be better investigated with the EVLA at high frequency or with the e-MERLIN array. 
\end{itemize}

\subsection{Jet velocity and orientation}
For 51 sources we can estimate the ratio between the jet and counter-jet brightness or provide a lower limit to the ratio .
If possible, we derived the core dominance CD, defined as the ratio between the observed arcsecond core power at 5 GHz P$_{core, observed}$ and the arcsecond core power P$_{core, estimated}$ estimated according to the Giovannini et al. 1994 correlation (GC) for the BCS sample (\cite{gio94}). 

In 8 sources the low core dominance suggests that the nuclear activity is now in a low activity state. The dominance of the extended emission implies a greater activity of the core in the past. However in these sources a parsec-scale core and even jets are present. In this scenario the nuclear activity may be in a low or high state but is not completely quiescent. This result is in agreement with the evidence that a few sources show evidence of a recurring or re-starting activity. This point can be better addressed when observations are available for the full sample so that we can discuss the time-scale of the recurring activity.

\subsection{FERMI DATA}

\begin{figure*}[t!]
\centering
\includegraphics[width=87mm]{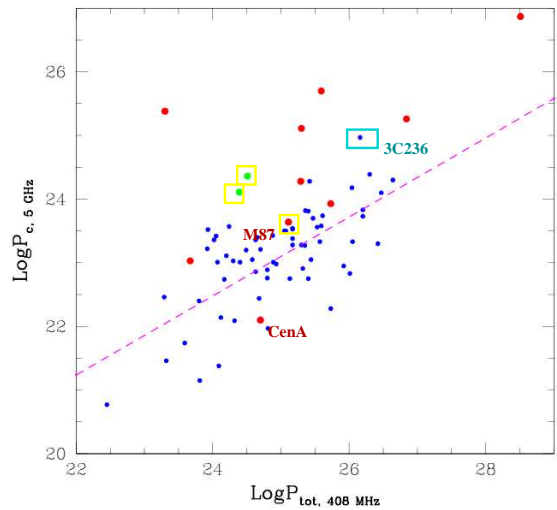}
\hfill
\includegraphics[width=87mm]{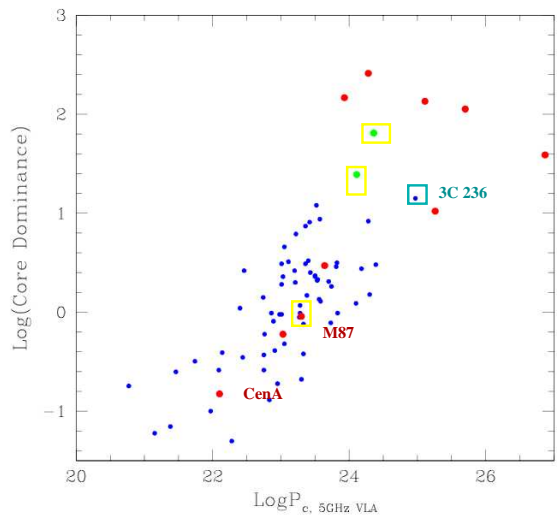}
\caption{ Top: {\bf LogP$_{core, 5 GHz}$ versus LogP$_{tot, 408 MHz}$}. BCS sources are blue points, the RGs in the 2FGL are the red ones, yellow circles indicate BCS that appears also in the 2FGL, green points correspond to BL Lacs BCS sources, azure square identify peculiar BCS radiogalaxy.  Bottom: {\bf LogP$_{core, 5 GHz}$ versus CD} } \label{PcPt_2012}
\end{figure*}

\begin{figure*}[t!]
\centering
\includegraphics[width=90mm]{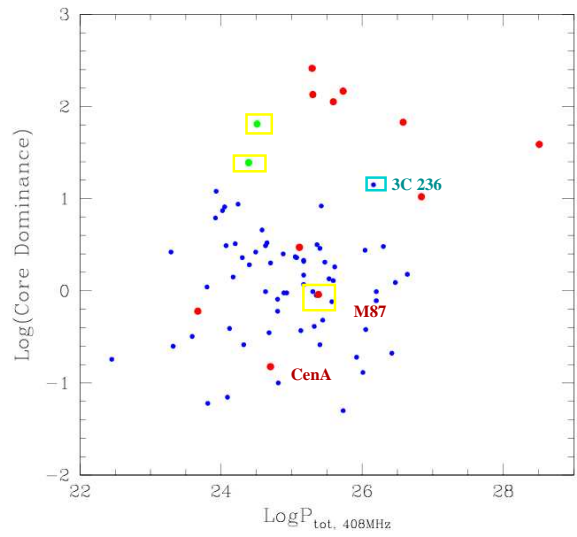}
\caption{{\bf Log P$_{tot, 408 MHz}$ versus CD}. For the symbols, refer to Fig. \ref{PcPt_2012}} \label{CdPt_2012}
\end{figure*}

We search in the LAT - 2year catalog the counterparts for our BCS sources: only the 2 BL Lacs (MKn 421 and Mkn 501) and M87 show gamma-ray emission.  To study a possibile relation between radio and gamma-ray emission for our complete sample of nearby objects, we investigated the following possible correlations:\\
a) P$_{c, 5 GHz}$ vs P$_{tot, 408 MHz}$ (Fig. \ref{PcPt_2012}),\\
b) CD (Core dominance) vs P$_{c, 5 GHz}$ (Fig. \ref{PcPt_2012}), and \\
c) CD vs P$_{tot, 408 MHz}$ (Fig. \ref{CdPt_2012}), \\
where P$_{c, 5 GHz}$ is the arcsecond power of the core at 5 GHz,  P$_{tot, 408 MHz}$ is total power of the source at 408 MHz. In the Fig.s: blue points are for the BCS radiogalaxies; green ones for the two BCS BL Lacs which are both 2LAT sources; red colour indicates the 2LAT radiogalaxies; yellow squares indicate BCS that appears also in the 2FGL; and azure square identify peculiar BCS radiogalaxy. In Fig. \ref{PcPt_2012}, the magenta dotted line represents the GC correlation.

Compared the BCS sources with radiogalaxies appearing in the 2FGL, our preliminary results are the followings:
\begin{itemize}
\item Among the 2FGL radiogalaxies, the majority of them are above the GC correlation, as expected  (\cite{gio94}) as they have high core dominance which indicates small angle of view $\theta$ of the object respect to the line of sight and boosted core.
\item In the P$_{core}$$-$CD plane, the 2FGL RGs seem to belong to a well defined region with high CD and high P$_{core}$.
\item M87 and Cen A are peculiar objects being below the GC :\\ I) M87 has small CD and large $\theta$ but it is detected by Fermi as a consequence of its proximity (z=0.04);\\ II) Cen A, despite its small CD, is a 2FGL sources as the Fermi emission is not only nuclear but it comes also from lobes.
\item 3C 236 is interesting being not in the 2FGL Catalogue: it has P$_{core}$, P$_{tot}$ and CD close to the 2FGL RGs but it is not a 2FGL source. In this case, the high CD is due to the restarted activities. It is like a young source. It could be detected by Fermi in the future.
\item In the P$_{tot}$$-$CD, the 2FGL RGs spread over P$_{tot}$, meaning independence of Fermi emission from P$_{tot}$, but they are segregated at high CD values that suggests the independence of the Fermi emission from P$_{tot, 408 MHz}$, but the presence of a relation with P$_{core}$, as beaming effects plus also intrinsically high P$_{core, 5 GHz}$ .
\item Gamma-ray emission from BCS sources with low CD could be detected by Fermi in the next future using stacking observations.
\end{itemize}

\begin{acknowledgments}
We thank the organizers of a very interesting meeting.
This work was supported by contributions of European Union, Valle D’Aosta Region and the Italian Minister for Work and Welfare.
\end{acknowledgments}

\end{document}